\documentclass[twocolumn,twoside,preprintnumbers,amsmath,amssymb,showkeys]{revtex4}
\usepackage{epsfig}
\usepackage{graphicx}

\newcommand{\beq}{\begin{equation}}
\newcommand{\eeq}{\end{equation}}
\newcommand{\bqa}{\begin{eqnarray}}
\newcommand{\eqa}{\end{eqnarray}}

\usepackage{fancyhdr}

\usepackage{pslatex}

\begin{document}

\title{The chromo-Weibel instability}

\author{Michael Strickland}

\affiliation{Frankfurt Institute for Advanced Studies \\
  Johann Wolfgang Goethe - Universit\"at Frankfurt \\
  Max-von-Laue-Stra\ss{}e~1 \\
  D-60438 Frankfurt am Main \\ 
  Germany }

\begin{abstract}

I discuss the physics of non-Abelian plasmas which are locally anisotropic in 
momentum space.  Such momentum-space anisotropies are generated by the rapid 
longitudinal expansion of the matter created in the first 1 fm/c of an 
ultrarelativistic heavy ion collision.  In contrast to locally isotropic plasmas 
anisotropic plasmas have a spectrum of soft unstable modes which are 
characterized by exponential growth of transverse chromo-magnetic/-electric 
fields at short times.  This instability is the QCD analogue of the Weibel 
instability of QED.  Parametrically the chromo-Weibel 
instability provides the fastest method for generation of soft background fields 
and dominates the short-time dynamics of the system.  The existence of the 
chromo-Weibel instability has been proven using diagrammatic methods, transport 
theory, and numerical solution of classical Yang-Mills fields.  I review the 
results obtained from each of these methods and discuss the numerical techniques 
which are being used to determine the late-time behavior of plasmas subject to a 
chromo-Weibel instability.

 \end{abstract}

\maketitle

\thispagestyle{fancy}

\setcounter{page}{1}

\section{Introduction}

With the ongoing ultrarelativistic heavy-ion collision experiments at the 
Relativistic Heavy-Ion Collider (RHIC) and planned the Large Hadron Collider 
(LHC) physicists hope to produce and study the properties of a thermalized 
quark-gluon plasma (QGP) which is expected to be formed when the temperature of 
nuclear matter is raised above its critical value, $T_c \sim 200$ MeV $\sim 
10^{12}$ K. Given the small size and short lifetime of the matter created in an 
ultrarelativistic heavy-ion collision this is not trivially accomplished.  One 
of the chief obstacles to thermalization in ultrarelativistic heavy-ion 
collisions is the rapid longitudinal expansion of the matter created in the 
central rapidity region. If the matter expands too quickly then there will not 
be sufficient time for its constituents to interact and thermalize. During the 
first 1 fm/c after the nuclear impact the longitudinal expansion causes the 
created matter to become much colder in the longitudinal direction than in the 
transverse directions~\cite{Baier:2000sb}, corresponding to $\langle p_L^2 
\rangle \ll \langle p_T^2\rangle$ in the local rest frame.  After this initial 
period of longitudinal cooling, the expansion slows and one can then ask what 
are the dominant mechanisms for driving the system towards an isotropic thermal 
QGP.

The evolution of the partonic matter created during a high-energy nuclear 
collision was among the questions which the ``bottom-up'' thermalization 
scenario~\cite{Baier:2000sb} attempted to answer. For the first time, it 
addressed the dynamics of soft modes (fields) with momenta much below $Q_s$ 
coupled to the hard modes (particles) with momenta on the order of $Q_s$ and 
above \cite{Mueller:2002kw,Iancu:2003xm,McLerran:2005kk}. However, it has 
emerged recently that one of the assumptions made in this model was not correct. 
The debate centers around the first stage of the bottom-up scenario 
($1 \ll Q_s \tau \ll \alpha^{-3/2}$) in 
which it was assumed that (a) collisions between the high-momentum (or hard) 
modes were the driving force behind isotropization and that (b) the low-momentum 
(or soft) fields acted only to screen the electric interaction. In doing so, the 
bottom-up scenario implicitly assumed that the underlying soft gauge modes 
behaved the same in an anisotropic plasma as in an isotropic one.  

However, to be self-consistent one must determine the collective modes which are 
relevant for an anisotropic plasma and use those.  In the case of gauge theories 
this turns out to be a qualitative rather than quantitative correction since in 
anisotropic QCD plasmas the most important collective mode corresponds to an 
instability to transverse chromo-magnetic field fluctuations 
\cite{Weibel:1959,Mrowczynski:1993qm,Mrowczynski:1994xv,Mrowczynski:1996vh}. 
This instability is the QCD analogue of the QED {\em Weibel 
instability}~\cite{Weibel:1959}.  Recent works have shown that the presence of 
the {\em chromo-Weibel instability} is generic for distributions which possess a 
momentum-space anisotropy 
\cite{Romatschke:2003ms,Arnold:2003rq,Romatschke:2004jh} and have obtained the 
full non-Abelian hard-loop (HL) action in the presence of an anisotropy 
\cite{Mrowczynski:2004kv}.  Another important development has been the 
demonstration that the chromo-Weibel instability also exists in solutions 
to pure classical Yang-Mills fields in an expanding geometry 
\cite{Romatschke:2005ag,Romatschke:2005pm,Romatschke:2006nk}.

Recently there have been significant advances in the understanding of non-Abelian 
soft-field dynamics in anisotropic plasmas within the HL 
framework~\cite{Rebhan:2004ur,Arnold:2005vb,Rebhan:2005re,Romatschke:2006wg}.  The HL 
framework is equivalent to the collisionless Vlasov theory of eikonalized hard 
particles, i.e.\ the particle trajectories are assumed to be unaffected (up to 
small-angle scatterings with $\theta \sim g$) by the induced background field. 
It is strictly applicable only when there is a large scale separation between 
the soft and hard momentum scales.  Even with these simplifying assumptions, HL 
dynamics for non-Abelian theories is complicated by the presence of 
non-linear gauge-field interactions.  

These non-linear interactions become important when the vector potential 
amplitude is on the order of $A_{\rm non-Abelian} \sim p_{\rm s}/g \sim f_h p_h$, 
where $p_h$ is the characteristic momentum of the hard particles, e.g. $p_h \sim 
Q_s$ for CGC initial conditions, $f_h$ is the angle-averaged occupancy at the 
hard scale, and $p_s$ is the characteristic soft momentum 
of the fields ($p_s \sim g f_h p_h$).  In QED there is no such complication and the 
fields grow exponentially until $A_{\rm Abelian} \sim p_h/g$ at which point the 
hard particles undergo large-angle deflections by the soft background field 
which rapidly isotropize the system.  In fact, in QED the Weibel instability is 
the fastest process driving plasma isotropization.  In QCD, however, the effect 
of the non-linear gauge self-interactions affects the system's dynamics 
primarily slowing down instability-driven particle isotropization.

To include the effects of gauge self-interactions numerical studies of the time 
evolution of the gauge-covariant HL equations of motion are required. Recent 
numerical studies of HL gauge dynamics for SU(2) gauge theory indicate that for 
{\em moderate} anisotropies the gauge field dynamics changes from exponential 
field growth indicative of a conventional Abelian plasma instability to linear 
growth when the vector potential amplitude reaches the non-Abelian scale, 
$A_{\rm non-Abelian} \sim p_{\rm h}$ \cite{Arnold:2005vb,Rebhan:2005re}. This 
linear growth regime is characterized by a turbulent cascade of the energy 
pumped into the soft modes by the instability to higher-momentum plasmon-like 
modes \cite{Arnold:2005ef,Arnold:2005qs}.  These results indicate that there is 
a fundamental difference between Abelian and non-Abelian plasma instabilities in 
the HL limit.

In addition to numerical studies in the HL limit there have been numerical results 
from the solution to the full non-linear Vlasov equations for anisotropic 
plasmas \cite{Dumitru:2005gp,Dumitru:2006pz}.  This approach can be shown to 
reproduce the HL effective action in the weak-field 
approximation~\cite{Kelly:1994ig,Kelly:1994dh,Blaizot:1999xk}; however, when 
solved fully the approach goes beyond the HL approximation since the full 
classical transport theory also reproduces some higher $n$-point vertices of the 
dimensionally reduced effective action for static gluons~\cite{Laine:2001my}. 
Numerical solution of the 3d Vlasov equations show that chromo-instabilities 
persist beyond the HL limit \cite{Dumitru:2006pz}.  Furthermore, 
the soft field spectrum obtained from full Vlasov simulations shows a cascade or 
``avalanche'' of energy deposited in the soft unstable modes in higher momentum 
modes similar to HL dynamics.

\section{Anisotropic Gluon Polarization Tensor}

In this section I consider a quark-gluon plasma with a parton distribution function
which is decomposed as~\cite{Romatschke:2003ms}
\bqa
f({\bf p}) \equiv 2 N_f \left[\, n_q({\bf p}) + n_{\bar q}({\bf p}) \,\right] 
+ 4 N_c\,n_g({\bf p}) \; ,
\label{distfncs}
\eqa
where $n_q$, $n_{\bar q}$, and $n_g$ are the distribution functions of quarks, anti-quarks, 
and gluons, respectively, and the numerical coefficients collect all appropriate
symmetry factors.
Using the result of Ref.~\cite{Romatschke:2003ms} the spacelike 
components of the HL gluon self-energy for gluons with soft 
momentum ($k \sim g\,p_{\rm hard}$) can be written as
\begin{equation}
\Pi^{i j}_{a b}(K) = - \frac{g^2}{2}\,\delta_{a b} \int \frac{d^3{\bf p}}{(2\pi)^3} v^{i}  \frac{\partial f({\bf p})}{\partial p^{l}}
\left( \delta^{j l}+\frac{v^{j} k^{l}}{K\cdot V + i \epsilon}\right) \; ,
\label{selfenergy2}
\end{equation}
%
where $K=(\omega,{\bf k})$, $V=(1,{\bf p}/p)$, and the parton distribution 
function $f({\bf p})$ is, up to an integrability requirement, completely 
arbitrary.  In what follows we will assume that $f({\bf p})$ can be obtained 
from an isotropic distribution function by the rescaling of only one direction 
in momentum space.  

In practice this means that, given any isotropic 
parton distribution function $f_{\rm iso}(p)$, we can construct an 
anisotropic version by changing the argument of the isotropic distribution 
function, 
$f({\bf p}) =  f_{\rm iso}\left(\sqrt{{\bf p}^2+\xi({\bf p}\cdot{\bf \hat n})^2}\right)$,
where ${\bf \hat n}$ is the direction of the anisotropy, and $\xi>-1$ is an adjustable 
anisotropy parameter with $\xi=0$ corresponding to the isotropic case. 
Here we will concentrate on $\xi>0$ which corresponds to a 
contraction of the distribution along the ${\bf \hat n}$ direction since this is 
the configuration relevant for heavy-ion collisions at early times, namely
two hot transverse directions and one cold longitudinal direction.

Making a change of variables 
in (\ref{selfenergy2}) it is possible to integrate out the $|p|$-dependence 
giving~\cite{Romatschke:2003ms}
\bqa
\Pi^{i j}_{a b}(\omega/k,\theta_n) &=& m_D^2 \, \delta_{a b} \int \frac{d \Omega}{4 \pi} v^{i}%
\frac{v^{l}+\xi({\bf v}\cdot\hat{\bf n}) \hat{n}^{l}}{%
(1+\xi({\bf v}\cdot\hat{\bf n})^2)^2} \nonumber \\
&& \hspace{1.5cm} \times
\left( \delta^{j l}+\frac{v^{j} k^{l}}{K\cdot V + i \epsilon}\right) ,
\eqa
where $\cos\theta_n \equiv \hat{\bf k}\cdot\hat{\bf n}$ and $m_D^2>0$. 
The isotropic Debye mass, $m_D$, depends on 
$f_{\rm iso}$ but is parametrically $m_D \sim g\, p_{\rm  hard}$.

The next task is to construct a tensor basis for the spacelike components of the 
gluon self-energy and propagator.  We therefore need a basis for symmetric 3-tensors 
which depend on a fixed anisotropy 3-vector $\hat{n}^{i}$ with 
$\hat{n}^2=1$.  This can be achieved with the following four component tensor 
basis: $A^{ij} = \delta^{ij}-k^{i}k^{j}/k^2$, $B^{ij} = k^{i}k^{j}/k^2$, $C^{ij} 
= \tilde{n}^{i} \tilde{n}^{j} / \tilde{n}^2$, and $D^{ij} = 
k^{i}\tilde{n}^{j}+k^{j}\tilde{n}^{i}$ with $\tilde{n}^{i}\equiv A^{ij} 
\hat{n}^{j}$. Using this basis we can decompose the self-energy into four 
structure functions $\alpha$, $\beta$, $\gamma$, and $\delta$ as ${\boldmath 
\Pi}= \alpha\,{\bf A} + \beta\,{\bf B} + \gamma\,{\bf C} + \delta\,{\bf D}$. 
Integral expressions for $\alpha$, $\beta$, $\gamma$, and $\delta$ can be found 
in Ref.~\cite{Romatschke:2003ms} and \cite{Romatschke:2004jh}.

\section{Collective Modes}

As shown in Ref.~\cite{Romatschke:2003ms} this tensor basis allows us to express the 
propagator in terms of the following three functions
\bqa
\Delta_\alpha^{-1}(K) &=& k^2 - \omega^2 + \alpha \; , \label{propfnc1} \nonumber \\
\Delta_\pm^{-1}(K) &=& \omega^2 - \Omega_\pm^2 \; , \nonumber
\label{propfnc2}
\eqa
where $ 2 \Omega_{\pm}^2 = \bar\Omega^2 \pm \left(\bar\Omega^4- 4 
((\alpha+\gamma+k^2)\beta-k^2\tilde n^2\delta^2) \right)^{1/2}$
and $\bar\Omega^2 = \alpha+\beta+\gamma+k^2$.  

Taking the static limit of these 
three propagators we find that there are three mass scales:  $m_\pm$ and $m_\alpha$. In 
the isotropic limit, $\xi\rightarrow0$, $m_\alpha^2=m_-^2=0$ and $m_+^2 = m_D^2$.  
However, for $\xi>0$ we find that $m_\alpha^2<0$ for 
all $\mid\!\!\theta_n\!\!\mid\,\neq\pi/2$ and $m_-^2<0$ for 
all $\mid\!\!\theta_n\!\!\mid\,\leq\pi/4$.  Note also that for $\xi>0$ both $m_\alpha^2$ and $m_-^2$
have there largest negative values at $\theta_n=0$ where they are equal.

\begin{figure}
\begin{center}
\includegraphics[width=8.7cm]{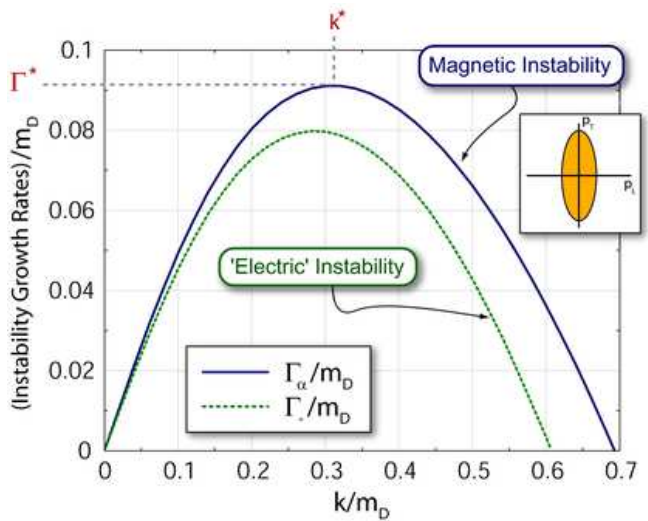}
\end{center}
\vspace{-7mm}
\caption{Instability growth rates as a function of wave number for $\xi=10$ and 
$\theta_n=\pi/8$.  Note that both growth rates vanish at $k=0$ and have a 
maximum $\Gamma^*\sim m_D/10$ at $k^*\sim m_D/3$.  The fact that they have a 
maximum means that at early times the system will be dominated by unstable modes 
with spatial frequency $\propto 1/k^*$ which grow at a rate $\Gamma^*$.}
\label{unstablemodesfig}
\end{figure}

The fact that for $\xi>0$ both $m_\alpha^2$ and $m_-^2$ can be negative 
indicates that the system is unstable to both magnetic and electric fluctuations 
with the fastest growing modes focused along the beam line ($\theta_n=0$).   In 
fact it can be shown that there are two purely imaginary solutions to each of the 
dispersion relations $\Delta_\alpha^{-1}(K)=0$ and $\Delta_- ^{-1}(K)=0$ with the 
solutions in the upper half plane corresponding to unstable modes.  We can 
determine the growth rate for these unstable modes by taking $\omega\rightarrow 
i\Gamma$ and then solving the resulting dispersion relations for $\Gamma(k)$.
Typical dispersion relations are shown in Fig.~\ref{unstablemodesfig}.

\section{Discretized Hard-Loop Dynamics}%

It is possible to go beyond an analysis of gluon polarization tensor to a
full effective field theory for the soft modes and then solve this numerically.  
The effective field 
theory for the soft modes that is generated by integrating out the hard plasma 
modes at one-loop order and in the approximation that the amplitudes of the soft 
gauge fields obey $A \ll |\mathbf p|/g$ is that of the gauge-covariant 
collisionless Boltzmann-Vlasov equations \cite{Blaizot:2001nr}. In equilibrium, the 
corresponding (nonlocal) effective action is the so-called hard-thermal-loop 
effective action which has a simple generalization to plasmas with anisotropic 
momentum distributions \cite{Mrowczynski:2004kv}. For the general non-equilibrium
situation the resulting equations of motion are
\begin{eqnarray}
D_\nu(A) F^{\nu\mu} &=& -g^2 \int {d^3p\over(2\pi)^3} {1\over2|\mathbf p|} \,p^\mu\, 
						 {\partial f(\mathbf p) \over \partial p^\beta} W^\beta(x;\mathbf v) \, , \nonumber \\
F_{\mu\nu}(A) v^\nu &=& \left[ v \cdot D(A) \right] W_\mu(x;\mathbf v) \, , 
\label{eom}
\end{eqnarray}
where $f$ is a weighted sum of the quark and gluon distribution functions 
\cite{Mrowczynski:2004kv} and $v^\mu\equiv p^\mu/|\mathbf p|=(1,\mathbf v)$.

These equations include all hard-loop resummed propagators and vertices and are 
implicitly gauge covariant.  At the expense of introducing a continuous set of 
auxiliary fields $W_\beta(x;\mathbf v)$ the effective field equations are also 
local.  These equations of motion are then discretized in space-time and 
${\mathbf v}$, and solved numerically.  The discretization in ${\mathbf v}$-space 
corresponds to including only a finite set of the auxiliary fields 
$W_\beta(x;\mathbf v_i)$ with $1 \leq i \leq N_W$. For details on the precise 
discretizations used see Refs.~\cite{Arnold:2005vb,Rebhan:2005re}.

\begin{figure}[t]
\vspace{4mm}
\centerline{
\includegraphics[width=8.7cm]{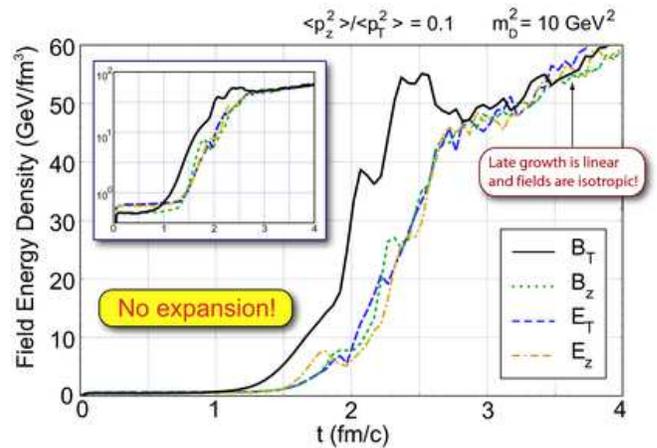}
}
\vspace{-2mm}
\caption{
Plot of typical energy densities observed in non-expanding three-dimensional 
hard-loop simulation of the soft-fields generated in an anisotropic plasma with 
$\xi=10$.  Shows transition from exponential growth with preference for 
transverse magnetic fields to linear isotropic growth.  Inset shows same data on 
logarithmic vertical axis.  To obtain physical units we have used $\alpha_s=0.3$ 
and $Q_s = 1.5\;{\rm GeV}$.
\label{3dHLfig}}
\end{figure}

\subsection{Discussion of Hard-Loop Results}

During the process of instability growth the soft gauge fields get the energy 
for their growth from the hard particles and, of course, the total energy is 
conserved.  In an Abelian plasma the energy deposited in soft fields grows 
exponentially until the energy in the soft fields is of the same order as the 
energy remaining in the hard particles at which point the back-reaction of the 
fields on the particle motion causes rapid isotropization.  As mentioned above 
in a non-Abelian plasma the situation is quite different and one must rely on 
numerical simulations due to the presence of strong gauge field self-
interactions. 

\begin{figure}[t]
\vspace{4mm}
\centerline{
\includegraphics[width=8.4cm]{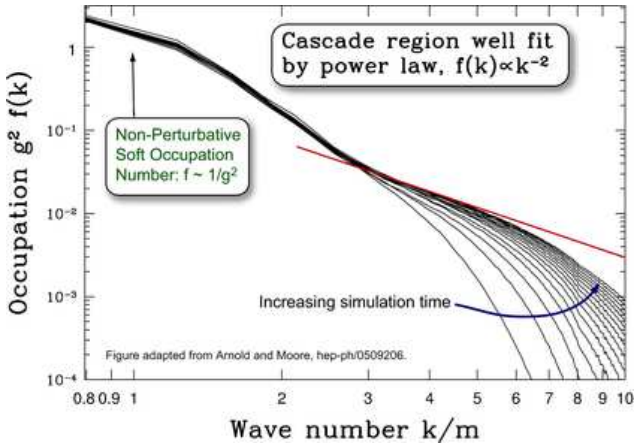}
}
\vspace{-2mm}
\caption{
Field mode spectrum for SU(2) runs showing saturation of soft field growth at $f \sim 1/g^2$
and an associated cascade of energy to the UV as the simulation time increases.  Figure 
adapted from Ref.~\cite{Arnold:2005ef}.
\label{cascadefig}}
\end{figure}

In Fig.~\ref{3dHLfig} I have plotted the time dependence of the chromo-magnetic/-electric 
energy densities obtained from a 3+1 dimensional from a 
typical HL simulation run initialized with ``weak'' random color noise with 
$\xi=10$ ~\cite{Rebhan:2005re}.  The inset shows the data on a logarithmic 
scale.  As can be seen from this figure at $t \simeq 2.5\;{\rm fm/c}$ there is a 
change from exponential to linear growth. Another interesting feature of the 
isotropic linear growth phase is that it exhibits a cascade of energy pumped 
into the unstable soft modes to higher energy plasmon like modes.  This is 
demonstrated in Fig.~\ref{cascadefig} which shows the soft gauge field spectrum 
as a function of momentum at different simulation times in the saturated linear 
regime along with the estimated scaling coefficient of the 
spectrum~\cite{Arnold:2005ef}. 

Note, in addition, Ref.~\cite{Arnold:2005ef} showed that there were non-perturbatively 
large chaotic Chern-Simons number fluctuations in the linear 
growth phase.  This should be contrasted to an Abelian theory or weak-field non-Abelian 
theory in which Chern-Simons number can fluctuate around zero but does 
not give large values when it is averaged over time. This is further indication of the 
non-perturbative physics associated with the chromo-Weibel instability.

From Fig.~\ref{3dHLfig} we can conclude that the chromo-Weibel instability will 
be less efficient at isotropizing a QCD plasma than the analogous Weibel 
instability seen in Abelian plasmas due to the slower than exponential growth at 
late times. On the positive side, from a theoretical perspective ``saturation'' 
at the soft scale implies that one can still apply the hard-loop effective 
theory self-consistently to understand the behavior of the system in the linear 
growth phase.
 
One caveat is that the latest published HL simulation results 
\cite{Arnold:2005vb,Rebhan:2005re} are for distributions with a finite 
${\mathcal O}(1\!\!\rightarrow\!\!10)$ anisotropy due to computational 
limitations and saturation seems to imply that for weak anisotropies field 
instabilities will not rapidly isotropize the hard particles.  This means, 
however, that due to the continued expansion of the system the anisotropy will 
increase.  It is therefore important to understand the behavior of the system 
for more extreme anisotropies. More importantly it is necessary to study the 
hard-loop dynamics in an expanding system.  Naively, one expects this to change 
the growth from $\exp(\tau)$ to $\exp(\sqrt\tau)$ at short times but there is no 
clear expectation of what will happen in the linear regime.  A significant 
advance in this regard has occurred recently for a U(1) gauge theory \cite{Romatschke:2006wg}.  
Work is underway to do the same for non-Abelian gauge theories.

\section{Wong-Yang-Mills equations}
\label{sec_WYMeqs}

It is also to possible to go beyond the hard-loop approximation and solve
instead the full classical transport equations in three dimensions \cite{Dumitru:2006pz}.
The Vlasov transport
equation for hard gluons with non-Abelian color charge $q^a$ in the
collisionless approximation are~\cite{Wong:1970fu,Heinz:1983nx},
\begin{equation}
 p^{\mu}[\partial_\mu - gq^aF^a_{\mu\nu}\partial^\nu_p
    - gf_{abc}A^b_\mu q^c\partial_{q^a}]f(x,p,q)=0~.   \label{Vlasov}
\end{equation}
Here, $f(t,\bf{x},\bf{p},q^a)$ denotes the single-particle phase space
distribution function.

The Vlasov equation is coupled self-consistently to the Yang-Mills
equation for the soft gluon fields,
\begin{equation}
 D_\mu F^{\mu\nu} = J^\nu = g \int \frac{d^3p}{(2\pi)^3} dq \,q\,
 v^\nu f(t,\bf{x},\bf{p},q)~, \label{YM}
\end{equation}
with $v^\mu\equiv(1,\bf{p}/p)$. These equations reproduce the
``hard thermal loop'' effective action near
equilibrium~\cite{Kelly:1994ig,Kelly:1994dh,Blaizot:1999xk}. However, 
the full classical transport
theory~(\ref{Vlasov},\ref{YM}) also reproduces some higher $n$-point
vertices of the dimensionally reduced effective action for
static gluons~\cite{Laine:2001my} beyond the hard-loop approximation. The
back-reaction of the long-wavelength fields on the hard particles
(``bending'' of their trajectories) is, of course, taken into account,
which is important for understanding particle dynamics in strong
fields.

Eq.~(\ref{Vlasov}) can be solved numerically by replacing
the continuous single-particle distribution $f(\bf{x},\bf{p},q)$
by a large number of test particles:
\bqa
 f(\bf{x},\bf{p},q) &=& \frac{1}{N_{\rm test}}\sum_i 
  \, \delta(\bf{x}-\bf{x}_i(t)) \nonumber \\
  && \hspace{6mm} \times (2\pi)^3 \delta(\bf{p}-\bf{p}_i(t)) \,
   \delta(q^a-q_i^a(t))~,  \label{TestPartAnsatz}
\eqa
where $\bf{x}_i(t)$, $\bf{p}_i(t)$ and ${q}_i^a(t)$ are the position, momentum, 
and color phase-space coordinates of an individual test particle and $N_{\rm 
test}$ denotes the number of test-particles per physical particle. The {\sl 
Ansatz}~(\ref{TestPartAnsatz}) leads to Wong's 
equations~\cite{Wong:1970fu,Heinz:1983nx}
\begin{eqnarray}
\frac{d\bf{x}_i}{dt} &=& \bf{v}_i \, ,\\ 
\frac{d\bf{p}_i}{dt} &=& g\,q_i^a \, 
							\left( \bf{E}^a + \bf{v}_i \times \bf{B}^a \right) \, ,\label{pdot}\\
\frac{d\bf{q}_i}{dt} &=& ig\, v^{\mu}_i \, [ A_\mu, \bf{q}_i] \,, \\ 
J^{a\,\nu} &=& \frac{g}{N_{\rm test}} \sum_i q_i^a \,v^\nu
\,\delta(\bf{x}-\bf{x}_i(t)) \, .
\end{eqnarray}
for the $i$-th test particle.  The time evolution of
the Yang-Mills field can be followed by the standard Hamiltonian
method~\cite{Ambjorn:1990pu} in $A^0=0$ gauge.  For details of the 
numerical implementation used see Ref.~\cite{Dumitru:2006pz}.

\begin{figure}[t]
\vspace{4mm}
$\;\;$\includegraphics[width=10.4cm]{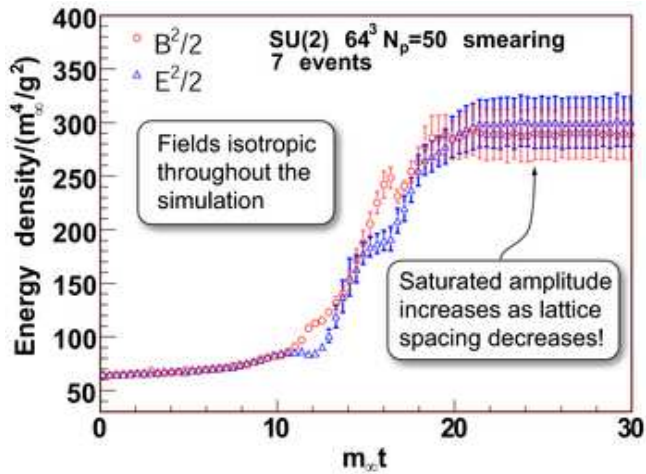}
\vspace{-5mm}
\caption{
Time evolution of the field energy densities for $SU(2)$
gauge group resulting from a highly anisotropic initial particle
momentum distribution.  Simulation parameters are $L=5$~fm, $p_{\rm
hard}=16$~GeV, $g^2\,n_g=10/$fm$^3$, $m_\infty=0.1$~GeV.}
\label{cpicfig}
\end{figure}

In Fig.~\ref{cpicfig} I present the results of a three-dimensional Wong-Yang-Mills 
(WYM) simulation published in Ref.~\cite{Dumitru:2006pz}.  The figure 
shows the time evolution of the field energy densities for $SU(2)$ gauge group 
resulting from a highly anisotropic initial particle momentum distribution.
The behavior shown in Fig.~\ref{cpicfig} indicates that the results obtained 
from the hard-loop simulations and direct numerical solution of the WYM 
equations are qualitatively similar in that both show that for non-Abelian 
gauge theories there is a saturation of the energy transferred to the soft modes 
by the gauge instability.  Although I don't show it here the corresponding 
Coulomb gauge fixed field spectra show that the field
saturation is accompanied by an ``avalanche'' of energy transferred to soft 
field modes to higher frequency field modes with saturation occurring when the 
hardest lattice modes are filled ~\cite{Dumitru:2006pz}.

A more thorough analytic understanding of this ultraviolet avalanche is lacking 
at this point in time although some advances in this regard have been made 
recently \cite{Mueller:2006up}.  Additionally, since within the numerical 
solution of the WYM equations the ultraviolet modes become populated rapidly 
this means that the effective theory which relies on a separation between hard 
(particle) and soft (field) scales breaks down.  This should motivate research 
into numerical methods which can be used to ``shuffle'' field modes to particles 
when their momentum becomes too large and vice-versa for hard particles.  Hopefully, 
using such methods it will be possible to simulate the non-equilibrium dynamics 
of anisotropic plasmas in a self-consistent numerical framework which treats 
particles and fields and the transmutation between these two types of degrees of 
freedom self-consistently.

\section{Outlook}

An important open question is whether quark-gluon plasma instabilities and/or 
the physics of anisotropic plasmas in general play an important phenomenological 
role at RHIC or LHC energies.  In this regard the recent papers of 
Refs.~\cite{Romatschke:2003vc,Romatschke:2004au,Schenke:2006fz,Romatschke:2006bb} 
provide theoretical frameworks which can be used to calculate the impact of 
anisotropic momentum-space distributions on observables such as jet shapes and 
the rapidity dependence of medium-produced photons. A concrete example of work 
in this direction is the recent calculation of photon production from an 
anisotropic QGP \cite{Schenke:2006aa}.  The results of that work suggest that it 
may be able possible to determine the time-dependent anisotropy of a QGP by 
measuring the rapidity dependence of high-energy medium photon production.


\bibliography{strickland}


\end{document}